\documentclass{article}

\begin{document}

\begin{titlepage}
\title{General Formula for Comparison of Clock Rates
  --- Applications to Cosmos and Solar System}
\author{Chongming Xu\thanks{To whom correspondence should be addressed
(cmxu@pmo.ac.cn, cmxu@njnu.edu.cn)}, \ Xuejun Wu \\
Purple Mountain Observatory, Nanjing 210008, China \\
Department of Physics, Nanjing Normal University, Nanjing 210097,
China  \and Erwin Br\"uning \\
School of Mathematical Sciences,
University of KwaZulu-Natal, \\
Durban 4000, South Africa}

\date{\today}

\maketitle

\begin{abstract}
In this paper we deduce a quite general formula which allows the
relation of clock rates at two different space time points to be
discussed. In the case of a perturbed Robertson-Walker metric, our
analysis leads to an equation for the comparison of clock rates at
different cosmic space time points, which includes the Hubble
redshift, the Doppler effect, the gravitational redshift and the
Rees-Sciama effects. In the case of the solar system, when the 2PN
metric is substituted into the general formula, the comparison of
the clock rates on both the earth and a space station could be
made. It might be useful for the discussion on the precise
measurements on future ACES and ASTROD.
\end{abstract}


\end{titlepage}

%%%%%%%%%%%%%%%%%%%%%%%%%%%%%%%%%%%%%%%%%%%%%%%%%%%%%%%%%%%%%%%%%%%%%%%

\setcounter{page}{2}

\section{Introduction}

Recently, atomic clocks with a time-keeping accuracy of the order
of $10^{-18}$ in fractional frequency have been considered
\cite{holz00,udem01}. Also a spatial experiment named ACES (Atomic
Clock Ensemble in Space) mission \cite{salo96,spal97} is scheduled
to be launched in near future by ESA (European Space Agency). The
purpose of ACES is to obtain an accuracy of order $10^{-16}$ in
fractional frequency. In such a situation ($10^{-16} - 10^{-18}$
level), 2PN (second Post Newtonian) approximate framework has to
be carried out before hand. Also, ASTROD (Astro-dynamical Space
Test of Relativity using Optical Devices) \cite{ni02,ni04} has
been proposed which is now under consideration by the Chinese
government. It is a three advanced drag-free spacecraft system,
and mature laser interferometric ranging will be used. The
accuracy of measuring $\gamma$ (about $10^{-9}$) and other
parameters will depend on the stability of the lasers or clocks.
In ASTROD the desired accuracy of the optical clock is about
$10^{-17}$. This plan also needs a 2PN level on the comparison of
clock rates and equations of motion for planets. The precision of
2PN level on the comparison of clock rates (or time transfer) has
been discussed in Ref. \cite{line02} by means of world function.
But as we know, the calculation of the world function is not easy
\cite{ponc04}. Also they discuss the frequency shift to the order
of $c^{-4}$ only in the field of an axisymmetric rotating body,
not in a general case. Therefore we deduce a general formula in a
different way. Our general formula can also be applied to cosmos
and easily extended to an even higher order (higher than 2PN
level). But here we have to emphasize that the purpose of our
general formula is to discuss the frequency shift (or variation of
the period) between two arbitrary space time points in a metric
theory. It is not to attempt to establish the synchronization of
the distant clocks, e.g. a grid of admissible radar 4-coordinates
\cite{alba03,alba05}, which is very useful for a GPS (Global
Positioning System) system of spacecrafts. It is limited only to
the clock rate, not the synchronization of the clock. It does not
involve the splitting of the space time.

We hope our general formula might be useful in ACES, ASTROD and
LISA (Laser Interferometer Space Antenna) \cite{lisa00} which is a
worldwide interested space mission. Our formula might be also
useful for the precise measurement of Hannay effect
\cite{hann85,spal04,spal05} (the Doppler shift variation is equal
to $3.7 \times 10^{-15}$ for a geostationary satellite). Hannay
effect is the classical counterpart of the quantum geometric phase
discovered by Berry \cite{berr84}. In future experiment, we have
to know about the higher precision of the modelling frequency
shift before hand. In this way, our formula may be useful for the
establishment of such a model.

On the other hand, one of the most basic experiments in physics is
the measurement of time, often at different positions in space.
Many formulae have been suggested for the comparison of clock
rates at different positions, based on certain simplifying
assumptions about which effects are dominant. We mention the most
prominent ones. The change of the clock rates can be related to:
the relativistic Doppler effect, the gravitational redshift, the
Hubble redshift, the Rees-Sciama effect and so on. Let us recall
these simple formulae on the comparison of clock rates in the
following.

In special relativity, the Doppler effect is shown by
\begin{equation}\label{1eq}
\frac{\Delta \tau_B}{\Delta \tau_A} = \frac{1- \frac{{\bf
  v}_{AB}}{c} \cdot \frac{\bf k}{|k|}}{\sqrt{1
  - \frac{v^2_{AB}}{c^2}}} \, ,
\end{equation}
where $\Delta \tau_A$ and $\Delta \tau_B$ are the proper period of
a light signal emitted by a source A and received by B
respectively, ${\bf v}_{AB}$ is the relative velocity from A to B,
$\bf k$ is the wave vector in flat space from A to B. The
relativistic Doppler effect (or transverse Doppler effect) $ 1/
\sqrt{(1 - \frac{v^2}{c^2})}$ was measured long time ago
\cite{ives38}. Recently the time dilation factor has been measured
with an accuracy of $2.2 \times 10^{-7}$ \cite{saat03}.

The gravitational redshift is given by
\begin{equation}\label{2eq}
\frac{\Delta \tau _B}{\Delta \tau _A} =
 \frac{\sqrt{-g_{00}(B)}}{\sqrt{-g_{00}(A)}} \, .
\end{equation}
In Newtonian limitation the metric tensor $g_{00}= -1 + 2w/c^2$,
then Eq.(\ref{2eq}) becomes
\[
\frac{\Delta \tau _B}{\Delta \tau _A} =
 1 - \frac{w(B)}{c^2} + \frac{w(A)}{c^2} + O(4) \, ,
\]
where $w$ is the Newtonian gravitational potential, $O(4)$ is the
abbreviation symbol for $O(c^{-4})$ as well as $O(n)$ for
$O(c^{-n})$ in the following formulae. The gravitational redshift
was measured with an accuracy of $7\times 10^{-5}$ \cite{vess80}.

The Hubble redshift is written as
\begin{equation}\label{3eq}
\frac{\Delta\tau_B}{\Delta\tau_A} = \frac{R(B)}{R(A)}\, ,
\end{equation}
where $\Delta\tau_B$ and $\Delta\tau_A$ are the period of
receiving and emitting signal, R is the cosmic scalar factor.

When photons pass freely through a region of voids or
supercluster, an additional redshift (or blue-shift) will be
created which has to be added to the normal Hubble redshift.
Because of the pioneer work by Rees $\&$ Sciama \cite{rees68},
this additional redshift due to the voids or superclusters is
called as Rees-Sciama effect. In their original paper they
modelled the voids or superclusters by a spherically symmetric
density over a Swiss cheese embedded in the Robertson-Walker
metric. The physical meaning of the effect is explained in detail,
but no derivation from the first principle is given. Later many
scientists \cite{blum92,tull92, mesz94} discussed the anisotropy
of CMBR(Cosmic Microwave Background Radiation) caused by
inhomogeneous Universe (Rees-Sciama effect). The original
(Ress-Sciama) spherical Swiss cheese model has been extended to an
integration of the time dependent Newtonian potential
\cite{tulu96}
\begin{equation}\label{4eq}
\frac{\Delta \tau_B - \Delta \tau_A}{\Delta \tau_A} = - w
(t_B,x^i_B)
 + w (t_A,x^i_A) + 2 \int _{A}^{B} \frac{\partial w}{\partial t}
 dt \, ,
\end{equation}
where $w$ is the Newtonian potential.

The physical conditions causing all these effects may be all
present. In early 90's the Hubble redshift, the gravitational
redshift, the Doppler effect and the Ress-Sciama effect have been
combined into one equation (see Eq.(6) in \cite{mart90}) in first
order approximation:
\begin{equation}\label{4eqA}
1+z = \frac{R(\tau_o)}{R(\tau_e)} \left\{ 1+\frac{5}{3}
  (\phi_e - \phi_o) + 2 \int^{\tau_o}_{\tau_e}
  d\tau {\bf l} \cdot \nabla \phi + {\bf n} \cdot
  ({\bf v}_e - {\bf v}_o) \right\} \, ,
\end{equation}
where $z$ is redshift, $\phi$ is gravitational potential, the
subscript $e(o)$ denote the emitting (observer) point , ${\bf
l}={\bf k}/ k_o$ ($k^\alpha$ is the tangent vector to the null
geodesic connecting the emitting point and observer), $\bf v$ is
3-velocity. Since all of terms are the level of the first order
approximation, the coupling terms do not exist. Martinez-Gonzalez
et al.\cite{mart90} do not deduce the equation (\ref{4eqA})
through an exact method, it is difficult for us to extend the
formula to higher order precision. At last, the coefficient $5/3$
may not be clear.

Accordingly, a comprehensive approach, starting from the first
principle, is needed in which the physical conditions for all
these effects are taken into account at the same time. Such an
approach should lead us to a synthetic formula which reflects all
these effects in a compact way and which should provide additional
information, due to possible interactions which could not be
incorporated in the isolated approaches for the individual
effects.

Normally the clock rates of A and B are represented by the proper
time interval $\Delta \tau_A $ and $\Delta \tau_B$ at position A
and B respectively. In general a comparison of the clock rates
between $\Delta \tau _A$ and $\Delta \tau _B$ by means of
differential coordinate time $\Delta t_A$ and $\Delta t_B$ in
global coordinates can be achieved. The relation of the coordinate
time between A and B is established by null geodesic line
(light-ray). The differential coordinate time on null geodesic
line can be easily calculated out \cite{wein72}
\begin{equation}\label{5eq}
c dt = \frac{- g_{0i}dx^i \pm \sqrt{ (g_{0i}g_{oj}
 -g_{00}g_{ij})dx^idx^j}}{g_{00}} \, .
\end{equation}
The minus and plus sign are taken in I and III quadrants (in x-t
coordinates) and II and IV quadrants respectively. Normally we
take the minus sign. Using these assumptions we can obtain a
general formula for the comparison of clock rates by means of some
integral.

In section II, we first time deduce a general formula by means of
``calculus of differences". In section III, substituting the
simplest perturbed Robertson-Walker metric into the general
formula, we obtain a formula for the comparison of clock rates at
different cosmic space time points, which includes the Hubble
redshift, the Doppler effect, the gravitational redshift and the
Rees-Sciama effects. In section IV, by using the 2PN metric in
multiple coordinates\cite{xu03}, the 2PN comparison of clock rates
on both the earth and a space station in the solar system is made,
being possibly useful for the precise measurement of ACES, ASTROD
and LISA in future. Some conclusion are offered in section V.

%%%%%%%%%%%%%%%%%%%%%%%%%%%%%%%%%%%%%%%%%%%%%%%%%%%%%%%%%%%%%%%%%%

\section{General formula}

In a global coordinates $(ct,x^i)$, a source A moves with a
velocity $v^i_A$ and a receiver B with a velocity $v^i_B$. The
clock rates in A and B are directly related with their own proper
time $\Delta \tau_A$ and $\Delta \tau_B$. To compare them, we need
to know the relation between the coordinate time interval $\Delta
t_A$ and $\Delta t_B$, because
\begin{equation}\label{A1}
\frac{\Delta \tau_A}{\Delta \tau_B} =
  \frac{\Delta \tau_A}{\Delta t_A}\frac{\Delta t_A}{\Delta t_B}
  \frac{\Delta t_B}{\Delta \tau_B} \, .
\end{equation}
Since  $-c^2 d \tau ^2 = ds^2 $, therefore if the velocity of a
standard clock (A or B) in the global coordinates is $\bf v$, we
have
\begin{eqnarray}
\Delta t_B &=& \frac{\Delta \tau _B}{\sqrt{-[g_{00}(B)+
 2g_{0i}(B)v^i_B /c + g_{ij}(B)v^i_B v^j_B /c^2]}}
 \, , \label{23eq}\\
\Delta t_A &=& \frac{\Delta \tau _A}{\sqrt{-[g_{00}(A)+
 2g_{0i}(A)v^i_A /c + g_{ij}(A)v^i_A v^j_A /c^2]}}
 \, ,\label{24eq}
\end{eqnarray}
where $g_{\mu\nu}$ ($g_{00}$, $g_{0i}$ and $g_{ij}$) are the
global metric. As abbreviation we introduce
\begin{eqnarray*}
G_A &=& -(g_{00}(A)+ 2g_{0i}(A)v^i_A /c
 + g_{ij}(A)v^i_A v^j_A /c^2)\, , \\
G_B &=& -(g_{00}(B)+ 2g_{0i}(B)v^i_B /c
 + g_{ij}(B)v^i_B v^j_B /c^2) \, .
\end{eqnarray*}
Here $\Delta t$, being an integrable coordinate time interval, has
unique meaning throughout space. Eq.(\ref{A1}) is known for long
time. In most discussions on the gravitational spectral shift,
people tend to consider a metric as independent of time
(stationary metric) and think points A and B are fixed in space,
then take $\Delta t_A = \Delta t_B$ for granted \cite{adle75}. One
of the main purpose of our paper is to calculate the relation
between $\Delta t_A$ and $ \Delta t_B$ by means of the ``calculus
of differences". Assuming that, at $t_{A_1}$ (coordinate time) a
source A emits a first pulse at position $A_1(x^i_{A_1})$, then a
receiver B receives the first pulse at position $B_1(x^i_{B_1})$
at time $t_{B_1}$. A second pulse is emitted from A at position
$A_2 (x^i_{A_2})$ at $t_{A_2}$, which is received by B at position
$B_2(x^i_{B_2})$ at time $t_{B_2}$. Two pulses transmit along the
null geodetic line 1 and 2 respectively. The emission time and
reception time are related as follows:
\begin{eqnarray}
t_{B_1} &=& t_{A_1} + \frac{1}{c} \int^{B_1}_{A_1}
 \frac{-g_{0i}dx^i - \sqrt{(g_{0i}g_{0j}- g_{00} g_{ij})
 dx^i dx^j}}{g_{00}} \, ,\label{6eq}\\
t_{B_2} &=& t_{A_2} + \frac{1}{c} \int^{B_2}_{A_2}
 \frac{-g_{0i}dx^i - \sqrt{(g_{0i}g_{0j}- g_{00} g_{ij})
 dx^i dx^j}}{g_{00}} \, .\label{7eq}
\end{eqnarray}
We define $d x^2 \equiv \delta _{ij}dx^idx^j$, the geometric
meaning of $dx$ is the spatial differential length of the line in
the flat space. Then the relation between the emission time and
reception time can be rewritten as
\begin{equation}\label{A2}
t_B = t_A + \frac{1}{c} \int^B_A
 \frac{-g_{0i}\frac{dx^i}{dx} - \sqrt{(g_{0i}g_{0j}- g_{00} g_{ij})
 \frac{dx^i}{dx} \frac{dx^j}{dx}}}{g_{00}} dx \, .
\end{equation}
In a weak field, $g_{0i}g_{0j}$ is a small quantity ($\sim O(6)$)
and the spatial conformal isotropic condition \cite{damo91,xu03}
is
\begin{equation}\label{A3}
g_{00}g_{ij}= - \delta _{ij} - \frac{q_{ij}}{c^4}
  + O(6)\, ,
\end{equation}
where $q_{ij}$ is a spatial anisotropic contribution in the second
order. Then Eq.(\ref{A2}) simplifies to
\begin{equation}\label{A4}
t_B = t_A - \frac{1}{c} \int^B_A \frac{1}{g_{00}}
 \left[ 1 + g_{0i}\frac{dx^i}{dx}
 + \frac{q_{ij}}{2c^4} \frac{dx^i}{dx} \frac{dx^j}{dx}
 \right] dx + O(6) \, .
\end{equation}
Briefly, we define
\begin{equation}\label{15eq}
F(t,x^i) \equiv \frac{-g_{0i}\frac{dx^i}{dx} -
\sqrt{(g_{0i}g_{0j}- g_{00} g_{ij})
 \frac{dx^i}{dx}\frac{ dx^j}{dx}}}{cg_{00}} \, ,
\end{equation}
Eq.(\ref{A2}) becomes $t_B=t_A + \int_A^B F(t,x^i)dx$. Considering
the time interval of two pulses as a period of the light frequency
(an atomic clock), it is very small (about $10^{-15}$ sec), during
which the source or the receiver move a distance only about a
magnitude of angstrom (\AA). Therefore in the following treatment
only the first order approximation (linear approximation) is taken
(all of higher order terms are neglected). From now on, we take
$A_1$ as $A$, $A_2$ as $A + \Delta A $ and as well as $B_1$ as
$B$, $B_2$ as $B + \Delta B $, $\Delta t_A = t_{A_2} - t_{A_1}$,
$\Delta t_B = t_{B_2} - t_{B_1}$. According to the ``calculus of
differences", we have
\begin{equation}\label{A3}
\Delta t_B = \Delta t_A + \Delta \left[ \int_A^B F(t,x^i)dx
 \right] \, ,
\end{equation}
where
\begin{equation}\label{16a}
\Delta \left[ \int_A^B F(t,x^i) dx \right] = \left.
 \int_{x(A +\Delta A)}^{x(B+\Delta B)} F(t,x^i) \right| _{\rm line2} dx
 - \left. \int_{x(A)}^{x(B)} F (t,x^i)\right| _{\rm line1} dx
 \, .
\end{equation}
If the metric is independent of time and points A and B are fixed
in space, Eq.(\ref{A3}) will lead to $\Delta t_A = \Delta t_B$
just as we mentioned earlier. Note that in Eq.(\ref{A3}) we are
dealing with finite differences.

The null geodetic line 1 and line 2 are very close, but different.
Therefore we can rewrite the integral along the line 2 as
\begin{equation}\label{16b}
\left. \int_{x(A +\Delta A)}^{x(B+\Delta B)} F(t,x^i)
 \right| _{\rm line2} dx = \left(
 \int_{x(B)}^{x(B+\Delta B)} + \int_{x(A)}^{x(B)}
 + \int_{x(A+\Delta A)}^{x(A)} \right)
 F(t+\Delta t, x+\Delta x) dx \, ,
\end{equation}
where $\Delta t$ and $\Delta x^i$ means the deviation from line 1
to line 2 (certainly $\Delta t$ and $\Delta x^i$ are the function
of space-time point). Substituting Eq.(\ref{16b}) in to
Eq.(\ref{16a}) and considering a linear approximation in the
``calculus of differences", the difference of the integral in
Eq.(\ref{A3}) can be divided into three parts:
\begin{equation}\label{A1a}
\Delta \int_A^B F dx = \int_{x(B)}^{x(B+\Delta B)} F dx
-\int_{x(A)}^{x(A+ \Delta A)} F dx
 + \int_{x(A)}^{x(B)} \Delta F dx \, ,
\end{equation}
where
\begin{equation}\label{17a}
\int_{x(A)}^{x(B)} \Delta F dx =
 \int_{x(A)}^{x(B)} \left( F(t+\Delta t, x^i + \Delta x^i)
 - F(t, x^i) \right) dx \, .
\end{equation}
$\Delta x(A)$ and $\Delta x(B)$ are given by
\begin{eqnarray}
\Delta x(A) &\equiv & x(A+\Delta A) - x(A)
 = \frac{{\bf k}_A}{|k_A|}\cdot
 \left. \frac{d{\bf x}}{dx} \right| _A \Delta x
 =\frac{{\bf k}_A}{|k_A|}\cdot {\bf v}_A\Delta t_A
   \, ,\label{16eq} \\
\Delta x(B) &\equiv & x(B+\Delta B) - x(B)
 = \frac{{\bf k}_B}{|k_B|}\cdot
 \left. \frac{d{\bf x}}{dx} \right| _B \Delta x
 =\frac{{\bf k}_B}{|k_B|}\cdot {\bf v}_B\Delta t_B
   \, .\label{17eq}
\end{eqnarray}
Here ${\bf k}_A$ is the wave vector at point A of the light signal
emitted from A and ${\bf k}_B$ is the value of the wave vector at
the received point B. In Eqs.(\ref{16eq}) and (\ref{17eq}) we have
replaced $\frac{dx^i}{dx} \Delta x$ by $\frac{dx^i}{dt} \Delta t$.
Because $\Delta t _A$  is a small quantity in the right hand side
of Eq.(\ref{16eq}), there is no difference between ${\bf k}(A_1)$
and ${\bf k}(A_2)$ at the level of linear approximation. We simply
take ${\bf k}_A$ to substitute ${\bf k}(A_1)$ and ${\bf k}(A_2)$,
as well as in Eq.(\ref{17eq}) ${\bf k}_B={\bf k}(B_1)={\bf
k}(B_2)$. The first and second terms in Eq.(\ref{A1a}) can be
written as
\begin{eqnarray}
\int_{x(B)}^{x(B)+x(\Delta B)} F dx &=& F(B)
  \frac{{\bf k}_B\cdot {\bf v}_B}{|k_B|}\Delta t_B
  \, ,\label{A5} \\
- \int_{x(A)}^{x(A)+x(\Delta A)} F dx &=& - F(A)
 \frac{{\bf k}_A\cdot {\bf v}_A}{|k_A|}\Delta t_A
  \, . \label{A6}
\end{eqnarray}
Eq.(\ref{17a}) is the difference of the integral between line 2
and line 1 when ${\bf v}_A = {\bf v}_B=0$ (i.e. the boundary of
integral is fixed), which can be expanded as
\begin{equation}\label{22eq}
\int_{x(A)}^{x(B)}
 \left( F(t+\Delta t,x^i+\Delta x^i) - F(t,x^i) \right) dx
  = \int_{x(A)}^{x(B)} \left( \frac{\partial F}{\partial t}
  \Delta t + \frac{\partial F}{\partial x^i} \Delta x^i \right) dx
  \, .
\end{equation}
As we know, if $F$ is independent of time, then
$\int_{x(A)}^{x(B)} \Delta F dx = \int_{x(A)}^{x(B)}
\frac{\partial F}{\partial x^i} \Delta x^i dx =0 $, since for
fixed boundaries the light ray is unique (no deviation). When $F$
is dependent on time, there are two curves. The second term $
\int_{x(A)}^{x(B)} \frac{\partial F}{\partial x^i} \Delta x^i dx
 $ caused by time dependent metric is a higher-order term
compared with $ \int_{x(A)}^{x(B)} \frac{\partial F}{\partial t}
\Delta t dx  $, i.e.
\begin{equation}\label{22Aeq}
\int_{x(A)}^{x(B)} \frac{\partial F}{\partial x^i} \Delta x^i dx
  \ll \int _{x(A)}^{x(B)}\frac{\partial F}{\partial t} \Delta t
  dx \, .
\end{equation}
In Ref. \cite{schn92} it is also mentioned that $
\int_{x(A)}^{x(B)} \frac{\partial F}{\partial x^i} \Delta x^i dx$
does not cause the change of frequency, even for cosmic distance.
Finally we substitute Eq.(\ref{A5}), (\ref{A6}), (\ref{22eq}) into
Eq.(\ref{A3}), and use Eq.(\ref{23eq}) and Eq.(\ref{24eq}) to
represent $\Delta t_A$ and $\Delta t_B$ by $\Delta \tau _A$ and
$\Delta \tau _B$. The general formula is obtained as
\begin{equation}\label{25eq}
\frac{\Delta \tau _B}{\Delta \tau _A} = \sqrt{\frac{G_B}{G_A}}
 \left( \frac{1 - F(A) \frac{{\bf k}_A \cdot
 {\bf v}_A}{|k_A|}}{1 - F(B) \frac{{\bf k}_B \cdot {\bf v}_B}{|k_B|}}
 \right)  + \frac{\sqrt{G_B}}{\Delta \tau _A \left( 1 - F(B)
 \frac{{\bf k}_B \cdot {\bf v}_B}{|k_B|} \right) }
 \int_{x(A)}^{x(B)} \left( \frac{\partial F}{\partial t}
  \Delta t + \frac{\partial F}{\partial x^i} \Delta x^i
  \right) dx \, .
\end{equation}
Eq.(\ref{25eq}) is our main result. Based on this equation we can
calculate the comparison of the clock rate between two arbitrary
points of space time. The equation is dependent on the metric and
the path of the null geodesic line. The metric can be solved from
the field equation (the general relativity certainly is a most
desired (but not the unique) theory) in different situations (the
energy momentum distribution and boundary conditions) while the
path of the null geodesic line can be obtained from null geodesic
equation. We do not care about the individual solution, even in
the following examples we only want to show how the formula is
efficient in deducing some rather complex formula (such as
Rees-Sciama effect) smoothly (see Section 3) or the coupling terms
 (see Section 4).

As an example, we consider the Doppler effect of a moving source
in Minkowski metric
\[
g_{\mu\nu}= \eta_{\mu\nu} = \left( \begin{array}{cc}
 -1 & 0 \\
 0 & \delta _{ij} \end{array} \right) \, ,
\]
since ${\bf v}_B=0$, $G_B=1$, $G_A= 1 - \frac{v_A^2}{c^2}$ and
$\frac{\partial F}{\partial t}= \frac{\partial F}{\partial x^i}
=0$, $F(A)= \frac{1}{c}$, so that
\begin{equation}\label{26eq}
\frac{\Delta \tau _B}{\Delta \tau _A} = \frac{1-\frac{\bf v}{c}
\cdot \frac{{\bf k}_A}{|k_A|}}{\sqrt{1-\frac{v_A^2}{c^2}}}\, .
\end{equation}
This is just the formula of the Doppler effect in the special
relativity (Eq.(\ref{1eq})).

The other simple example is the gravitational redshift.
Considering a static gravitational field (e.g. Schwarzschild
metric), in which both source and receiver without moving (${\bf
v}_A = {\bf v}_B =0$), the general form then becomes
\begin{equation}\label{27eq}
\frac{\Delta \tau_B}{\Delta \tau_A} =
 \sqrt{\frac{-g_{00}(B)}{-g_{00}(A)}}
 \simeq 1 - \frac{w(B)}{c^2} + \frac{w(A)}{c^2} \, ,
\end{equation}
where the last step of above equation is the Newtonian limitation.
Eq.(\ref{27eq}) is just the equation shown in normal textbooks of
gravity (see Eq.(\ref{2eq})).

Naturally, for an explicit evaluation of Equation (\ref{25eq}),
the light-ray trajectory should be known. If we take $dl=cdt$ as
the arc length of light-ray trajectory, then we get\cite{gong04}
\begin{equation}\label{9eq}
\frac{dx^i}{dl}= n^i + \alpha_1 \frac{dx^i_{1p}}{dl}
 + \alpha_2 \frac{dx^i_{2p}}{dl} + \cdots \, ,
\end{equation}
where $n^i$ is a constant unit vector, parameters $\alpha_1$ and
$\alpha_2$ $\cdots$ are equal to zero or 1 dependent on which
approximate level we consider. For 1PN level ($\alpha_1=1$,
$\alpha_i=0$ when $i\neq 1$), we have \cite{will93}
\begin{eqnarray}
&& {\bf n} \cdot \frac{d {\bf x}_{1p}}{dt} = -\frac{2 w}{c} \, ,
  \label{10eq}\\
&& \frac{d^2 {\bf x}_{1p}}{dt^2} = 2 \left[ \nabla w
  - 2 {\bf n}( {\bf n} \cdot \nabla w ) \right] \, ,
  \label{11eq}
\end{eqnarray}
For 2PN level ($\alpha_1=\alpha_2=1$, $\alpha_i=0$ when $i\geq
3$), by means of iterative method we deduce
\begin{eqnarray}
&& {\bf n}\cdot\frac{d{\bf x}_{2p}}{dt} = \frac{4w_in^i}{c^2}
 -\frac{1}{2c}\left|\frac{d{\bf x}_{1p}}{dt}\right|^2
 + \frac{4w^2}{c^3}- \frac{q_{ij}}{2c^3}n^in^j \, ,\label{12eq}\\
&& \frac{d^2x^i}{dt^2} = w_{,i} \left( 2 - \frac{4w}{c^2}
 + \frac{2{\bf n}}{c} \cdot \frac{d{\bf x}_{1p}}{dt} \right) \nonumber \\
 && \hskip8mm - 2\left( cn^i + \frac{dx^i_{1p}}{dt}\right)
  \left( \frac{2w_{,j}n^j}{c} + \frac{w_{,t}}{c^2}
  -\frac{2w_{(j,k)}}{c^2}n^jn^k
  +\frac{2 w_{,j}}{c^2}\frac{dx^j_{1p}}{dt} \right)
  \nonumber \\
 && \hskip8mm  + \frac{8w_{[i,j]}n^j}{c} - \frac{q_{ij,k}}{c^2}n^jn^k
  + \frac{q_{jk,i}}{2c^2}n^jn^k \, ,\label{13eq}
\end{eqnarray}
where $w$, $w^i$ and $q_{ij}$ are defined in \cite{damo91,xu03}.
Hereafter we use the abbreviation DSX for ``the papers written by
 Damour, Soffel and Xu" \cite{damo91,damo92,damo93,damo94}.
 In principle, if we knew the metric
to N-PN level, we could solve $\frac{dx^i}{dl}$ (or
$\frac{dx^i}{dt}$) up to such a N approximate level. Therefore
$\frac{dx^i}{dl}$ can be taken as a known function.

In cosmological problems, $dx^i$ has to be replaced by $R(t) dx^i$
where $R(t)$ is the cosmic scalar factor (or the expanding factor)
\cite{bert95}, $x^i$ is a comoving coordinate now. Then
Eq.(\ref{9eq}) has to be written as
\begin{equation}\label{14eq}
R(t)\frac{dx^i}{dl}= n^i + \alpha_1 R(t)\frac{dx^i_{1p}}{dl}
 + \alpha_2 R(t)\frac{dx^i_{2p}}{dl} + \cdots \, .
\end{equation}
Usually, in most problems of cosmology, the deviation from the
leading term would not be considered. The relation between $dx$
and $dl$ can be simply obtained from Eq.(\ref{9eq}) for
$\alpha_1=\alpha_2 =1$
\begin{equation}\label{A7}
\left( \frac{dx}{dl} \right) ^2 =
 1 + 2 {\bf n} \cdot \frac{d{\bf x}_{1p}}{dl}
 + 2 {\bf n} \cdot \frac{d{\bf x}_{2p}}{dl}
 + \frac{d{\bf x}_{1p}}{dl} \cdot \frac{d{\bf x}_{2p}}{dl}
 + O(6) \, .
\end{equation}
Now we shall discuss two applied examples in the following
section.

%%%%%%%%%%%%%%%%%%%%%%%%%%%%%%%%%%%%%%%%%%%%%%%%%%%%%%%%%%%%%%%%%

\section{Application in Cosmos with perturbed R-W Metric}

First we recall the unperturbed Robertson-Walker metric
\begin{equation}\label{28eq}
ds^2 = -c^2 dt^2 + \frac{R(t)^2 \delta_{ij}dx^idx^j}{\left( 1+
\frac{k}{4}r^2 \right)^2} \, ,
\end{equation}
where $R(t)$  is the cosmic scalar factor, $k=-1, \, 0, \, +1$ is
corresponding to the open, flat and closed universe respectively.
$R(t)$ has the dimension of length and $dx^i$ is dimensionless. As
we already know $R(t)$ is model dependent. In cosmology, by using
Eq.(\ref{28eq}) the light ray $ds^2 =0$ yields
\begin{equation}\label{29eq}
\frac{dt}{R(t)}
 = \frac{\sqrt{\delta _{ij}dx^idx^j}}{1+\frac{k}{4}r^2}
  \, .
\end{equation}
The right side of Eq.(\ref{29eq}) is independent of time,
therefore
\begin{equation}\label{30eq}
\frac{dt}{R(t)} = \frac{dt_A}{R(t_A)}
 = \frac{dt_B}{R(t_B)} \, .
\end{equation}
Since usually we do not consider the local gravitational redshift
and the Doppler effect in the problem of cosmological expansion,
we then have $\Delta t_A = \Delta \tau_A$ and $\Delta t_B = \Delta
\tau_B$, and thus the formula for the Hubble redshift is
\begin{equation}\label{31eq}
\frac{\Delta \tau_A}{\Delta \tau_B} = \frac{\Delta t_A}{\Delta
 t_B} = \frac{R(t_A)}{R(t_B)} \, .
\end{equation}
The results of Eq.(\ref{31eq}) can also be deduced directly from
the general formula Eq.(\ref{25eq}), if we take
$v_A=v_B=w_A=w_B=0$.

Next, we consider a linearly simplest perturbed Robertson-Walker
metric of the form
\begin{equation}\label{32eq}
ds^2 = -c^2 \left( 1 - \frac{2w}{c^2} \right) dt^2 + \left( 1 +
  \frac{2w}{c^2} \right) \frac{R^2 \delta_{ij}dx^idx^j}{\left( 1+
  \frac{k}{4}r^2 \right)^2} \, ,
\end{equation}
where the gravitational potential $w=w(t,x^i)$ is assumed to be a
small quantity. Certainly, we might choose more general
perturbation, e.g. Sachs-Wolfe integrated effect \cite{sach67} or
the gauge invariant perturbation \cite{bard80,elli89}. In fact if
you know any kind of metric, just substituting it into the general
formula (Eq.(\ref{25eq})) to get the comparison of clock rates
between any two arbitrary points. If we use some complex
perturbation, the formula on the comparison of clock rates would
be too long to see our extension clearly. By the way, the
application of our general formula in discussing CMBR \cite{anni}
might be very attractive for us in the future. For the moment we
will concentrate on solving the comparison of clock rates between
two arbitrary points. Here we consider the Doppler effect caused
only by the motion of the source, then ${\bf v}_B=0$ (also
possible ${\bf v}_A=0$, then ${\bf v}_B \neq 0 $). The velocity of
the source A is
\begin{equation}\label{33eq}
v^i_A = R(t_A) \frac{dx^i_A}{dt} \, .
\end{equation}
$F(A)$, $G_A$, and $G_B$ can be calculated as follows
\begin{equation}\label{34eq}
F(A) = \frac{ \left( 1 + \frac{2w}{c^2} \right) \sqrt{ R^2
 \delta_{ij}\frac{dx^i}{dx}\frac{dx^j}{dx}}}{ c \left( 1+
 \frac{kr^2}{4} \right)} = \frac{R(t_A)}{c} + O(3) \, ,
\end{equation}
where we have neglected all of higher order terms and consider
$\delta _{ij}n^in^j=1$ and $kr^2$ as higher order term also. The
quantities $G_A$ and $G_B$ read as
\begin{eqnarray}
\sqrt{G_B} &=& \sqrt{-g_{00}(B)}
  = 1 - \frac{1}{c^2} w(t_B, x_B^i) \, ,\label{35eq} \\
\sqrt{G_A} &=& \sqrt{-g_{00} -2g_{0i} \frac{dx^i}{dt} -
  g_{ij}\frac{dx^i}{dt} \frac{dx^j}{dt}}
  = 1 - \frac{1}{c^2} w(t_A, x_A^i) - \frac{1}{c^2} v^2_A
  \, ,\label{36eq} \\
\sqrt{\frac{G_B}{G_A}} &=& 1 + \frac{1}{c^2} w(t_A, x_A^i)
  - \frac{1}{c^2} w(t_B, x_B^i) + \frac{v^2_A}{c^2}
  \, ,\label{37eq}
\end{eqnarray}
Now we calculate the last term of Eq.(\ref{25eq}). We should pay
attention to that, in Eqs.(\ref{35eq}) and (\ref{36eq}) we use the
same symbol $w$ to represent the gravitational potential at point
A and point B, but really their form maybe very different (along
the right-ray a potential may caused by supercluster or the sun),
normally we can not use the same function to describe the
potential in different points A and B. In the last step we
consider the integral in Eq.(\ref{25eq})
\begin{equation}\label{38eq}
I \equiv \int_{x(A)}^{x(B)} \frac{\partial F}{\partial t} \Delta t
   dx \, ,
\end{equation}
where we have omit the term of $\int_{x(A)}^{x(B)} \frac{\partial
F}{\partial x^i} \Delta x^i dx $ because of Eq.(\ref{22Aeq}).
Therefore we have
\begin{equation}\label{39eq}
I = \int_{x(A)}^{x(B)} \frac{\partial }{\partial t} \left[
   R(t)\left( 1+\frac{2w}{c^2}\right) \right] \frac{ \Delta t
  \sqrt{ \delta_{ij} dx^i dx^j}}{ c \left( 1+\frac{k}{4}
  r^2 \right)} \, .
\end{equation}
In fact we know the relation between $\Delta t_A$ and $\Delta
t_B$, therefore at any time t, in first order of approximation
from Eq.(\ref{30eq}) we have
\begin{equation}\label{40eq}
\Delta t = \frac{R(t)}{R(t_A)} \Delta t_A \, .
\end{equation}
Eq.(\ref{39eq}) becomes
\begin{equation}\label{41eq}
I = \int_{x(A)}^{x(B)} \frac{R(t) \Delta t_A}{R(t_A)}
  \left( \dot{R}(t) + \dot{R}(t)\frac{2w}{c^2}
  + R(t) \frac{\partial (2w/c^2)}{\partial t} \right)
  \frac{\sqrt{\delta_{ij}dx^idx^j}}{c \left(
  1+ \frac{k}{4}r^2 \right) } \, .
\end{equation}
Considering null geodetic line, Eq.(\ref{32eq}) yields
\begin{equation}\label{42eq}
\left( 1 - \frac{w}{c^2} \right) dt = \pm
  \left( 1 + \frac{w}{c^2} \right)
  \frac{R(t) \sqrt{\delta_{ij}dx^idx^j}}{c\left(1 +
  \frac{k}{4}r^2\right)} \, ,
\end{equation}
which we use to evaluate the integral $I$ and get
\begin{equation}\label{43eq}
I = \frac{\Delta t_A}{R(t_A)} \int_{x(A)}^{x(B)} \left( \dot{R}(t)
 +2\frac{R(t)}{c^2}\frac{\partial w}{\partial t} \right) dt
  = \Delta t_A \left[ \left( \frac{R(t_B)-R(t_A)}{R(t_A)} \right)
  + \frac{2}{c^2 R(t_A)} \int_{x(A)}^{x(B)} R(t)
  \frac{\partial w}{\partial t} dt \right] \, .
\end{equation}
The second term of Eq.(\ref{25eq}) then becomes
\begin{equation}\label{44eq}
\sqrt{ \frac{G_B}{G_A}}\left( \frac{R(t_B)-R(t_A)}{R(t_A)}
  + \frac{2}{c^2 R(t_A)} \int_{x(A)}^{x(B)} R(t)
  \frac{\partial w}{\partial t} dt \right) \, .
\end{equation}
Substituting Eqs.(\ref{34eq}), (\ref{37eq}) and (\ref{44eq}) into
Eq.(\ref{25eq}) (pay attention to that, in cosmology the term $
\frac{\bf k}{|k|} \cdot {\bf v}$ should be replaced by $ \frac{\bf
k}{|k|} \cdot \frac{\bf v}{R}$ in Eq.(\ref{16eq}), (\ref{17eq}),
and Eq.(\ref{25eq})). we finally obtain
\begin{equation}\label{45eq}
\frac{\Delta \tau_B}{\Delta \tau_A} = \left[ 1 + \frac{1}{c^2}
 \left( w(t_A,x^i_A) - w(t_B,x^i_B) \right) +\frac{v^2_A}{c^2}
 \right] \left\{ \frac{R(t_B)}{R(t_A)}
 - \frac{{\bf k}_A \cdot {\bf v}_A}{c|k_A|}
 + \frac{2}{c^2R(t_A)} \int_{x(A)}^{x(B)} R(t) \frac{\partial w}{\partial t}
 dt \right\} \, ,
\end{equation}
where $\frac{1}{c^2}\left( w(t_A,x^i_A)- w(t_B,x^i_B)\right)$ is
the contribution from the normal gravitational redshift;
$\frac{{\bf k}_A \cdot {\bf v}_A}{c|k_A|}$ and $\frac{v^2_A}{c^2}$
are the Doppler effect and transverse Doppler effect (or
relativistic Doppler effect) respectively; $R(t_B)/R(t_A)$ just
contributes to Hubble redshift; the last term is related to
Rees-Sciama effect. Eq.(\ref{45eq}) can be compared with
Eq.(\ref{4eq}) if we put ${\bf v}_A=0$ and $R(t)=R(t_A)=R(t_B)$,
then
\begin{equation}\label{46eq}
\frac{\Delta \tau_B}{\Delta \tau_A} = 1 + \frac{1}{c^2}
 \left( w(t_A,x^i_A) - w(t_b,x^i_B) \right)
 + \frac{2}{c^2} \int_{x(A)}^{x(B)} \frac{\partial w}{\partial t}
 dt \, ,
\end{equation}
where $w$ is the same as $u$ in Ref. \cite{tulu96} where $c=1$
units is taken, and their results totally agree with ours. We
thought that our general formula allows to derive the
Birkinshaw-Gull effect \cite{birk83,birk89} too, if we use
suitable perturbation functions for $w(t,x^i)$ and $w_j(t,x^i)$.
This will be  discussed in another paper.

%%%%%%%%%%%%%%%%%%%%%%%%%%%%%%%%%%%%%%%%%%%%%%%%%%%%%%%%%%%%%%%%%%%%%%%

\section{Application in Solar System with DSX metric}

Our formula (\ref{25eq}) is of substantial generality and
accordingly allows a great number of fruitful applications. Some
of these have been given in the previous sections. However a
important application probably is to the solar system since there
we have a chance to do measurement and  thus by comparison with
measurements our formula can be tested. In this section, we
provide a fundamental formula with the precision $O(4)$, but we
have not expanded the potential and vector potential by means of
multiple moments that has to be done in the practical problems.

In near future high precision measurement will be done up to 2PN
level as we mentioned before, thus allowing coupling terms (i.e.,
the terms connecting the gravitational redshift, the Doppler
redshift and so on) to be measured. Our scheme (the general form
Eq. (\ref{25eq})) offers the possibility for this if an
appropriate assumptions about the metric are used. Accordingly, in
this section we start from DSX formalism
\cite{damo91,damo92,damo93} and its extension \cite{xu03} and
evaluate formula (\ref{25eq}) for this metric. This extended DSX
metric is described by
\begin{eqnarray}
&& g_{00} = -\exp{\left( - \frac{2w}{c^2} \right) } +O(6)
  \, , \label{47eq} \\
&& g_{0i} = - \frac{4w_i}{c^3} + O(5)\, , \label{48eq} \\
&& g_{ij} = \delta _{ij}\exp{\left( \frac{2w}{c^2} \right) }
  + \frac{q_{ij}}{c^4}+O(6) \, , \label{49eq} \\
&& g_{ij}g_{00} = - \delta _{ij}
  - \frac{q_{ij}}{c^4}+O(6) \, . \label{50eq}
\end{eqnarray}
In fact, in the following calculation, $q_{ij}$ appear only in the
function $F(x^i,t)$, but in the final 2PN formula of clock rates
(to see Eq.(\ref{63eq})) $q_{ij}$ does not exist which agree with
the result in Ref. \cite{line02}. Substituting
Eq.(\ref{47eq})-(\ref{50eq}) into Eq.(\ref{25eq}), calculating all
of components, we can get a general formula for the comparison of
clock rates at 2PN level.

We begin by evaluating the terms $G_A$ and $G_B$ for the extended
DSX metric:
\begin{eqnarray}
G_A &=& - g_{00}(A) - 2g_{0i}(A)\frac{v^i_A}{c}
  - g_{ij}(A) \frac{v^i_Av^j_A}{c^2} \nonumber \\
  &=& 1 - \frac{2w(A)}{c^2} + \frac{2w^2(A)}{c^4}
  + \frac{8w_i(A)v^i_A}{c^4} - \frac{v^2_A}{c^2}
  - \frac{2w(A)v^2_A}{c^4}+O(6) \, , \label{51eq} \\
G_B &=& 1 - \frac{2w(B)}{c^2} + \frac{2w^2(B)}{c^4}
  + \frac{8w_i(B)v^i_B}{c^4} - \frac{v^2_B}{c^2}
  - \frac{2w(B)v^2_B}{c^4} + O(6) \, . \label{52eq}
\end{eqnarray}
Next we consider the terms $F(A){\bf k}_A \cdot {\bf v}_A/|k_A|$
and $F(B){\bf k}_B \cdot {\bf v}_B/|k_B|$. Since ${\bf k}_A \cdot
{\bf v}_A/|k_A|$ and ${\bf k}_B \cdot {\bf v}_B/|k_B|$ are first
order already, so $F(A)$ and $F(B)$ need to be calculated up to
$c^{-5}$ level. From Eq.(\ref{15eq}) we get
\begin{equation}\label{53eq}
F(A) = \left. \frac{-g_{0i}\frac{dx^i}{dx} -
  \sqrt{(g_{0i}g_{0j}-g_{00}g_{ij})
  \frac{dx^i}{dx}\frac{dx^j}{dx}}}{cg_{00}} \right| _A
  = \left. \frac{-g_{0i}\frac{dx^i}{dx} - \sqrt{-g_{00}g_{ij}
  \frac{dx^i}{dx}\frac{dx^j}{dx}}}{cg_{00}} \right|_A \, ,
\end{equation}
where we have neglected $g_{0i}g_{0j}(\sim O(6))$. By using
Eq.(\ref{47eq}--\ref{50eq}), Eq.(\ref{53eq}) then becomes
\begin{equation}\label{54eq}
F(A) = \frac{1}{c} \left\{ 1 + \frac{2w(A)}{c^2} -
  \frac{4w_i(A)}{c^3} \left. \frac{dx^i}{dx} \right|_A
  + \frac{2w^2(A)}{c^4} + \frac{1}{2c^4} q_{ij}(A)
  \left. \left( \frac{dx^i}{dx}  \frac{dx^j}{dx}
  \right) \right| _A \right\} + O(6) \, ,
\end{equation}
and similarly
\begin{equation}\label{54eq2}
F(B) = \frac{1}{c} \left\{ 1 + \frac{2w(B)}{c^2} -
  \frac{4w_i(B)}{c^3} \left. \frac{dx^i}{dx} \right|_B
  + \frac{2w^2(B)}{c^4} + \frac{1}{2c^4} q_{ij}(B)
  \left. \left( \frac{dx^i}{dx} \frac{dx^j}{dx}
  \right) \right| _B \right\} + O(6) \, .
\end{equation}
At last, we consider the integral (the second term) in
Eq.(\ref{25eq})
\begin{equation}\label{55eq}
\int_A^B \left( \frac{\partial F}{\partial t} \Delta t
  + \frac{\partial F}{\partial x^i} \Delta x^i
  \right) dx \, ,
\end{equation}
where because of Eq.(\ref{22Aeq}) we will omit the second term in
above equation. Accordingly instead of (\ref{55eq}) we consider
\begin{equation}\label{56eq}
\int_A^B \frac{\partial F}{\partial t} \Delta t dx \, .
\end{equation}
As we know $ \Delta t_A$ at A and $\Delta t_B$ at B, we could
calculate $\Delta t$ at an arbitrary point between A and B
approximately. Then the definite integral (\ref{56eq}) can be
evaluated by the median method, i.e., %%
\begin{equation}\label{57eq}
\int_A^B \frac{\partial F}{\partial t} \Delta t dx
  = \Delta \bar{t} \int_A^B \frac{\partial F}{\partial t} dx \, ,
\end{equation}
where $\Delta \bar{t}$ is the median value, for which we introduce
a parameter $\eta$:
\begin{equation}\label{58eq}
\Delta \bar{t} = \eta \Delta t_A \, .
\end{equation}
$\eta$ is a value closed to $1$. For the term $\frac{\partial
F}{\partial t}$ one finds
\begin{equation}\label{A8}
\frac{\partial F}{\partial t} = \frac{1}{c} \left(
  \frac{2}{c^2}\frac{\partial w}{\partial t}
  - \frac{4}{c^3}\frac{\partial w_i}{\partial t}\frac{dx^i}{dx}
  + \frac{4w}{c^4}\frac{\partial w}{\partial t}
  + \frac{1}{2c^4}\frac{\partial q_{ij}}{\partial t}
  \frac{dx^i}{dx}\frac{dx^j}{dx} \right) + O(6) \, .
\end{equation}
In the solar system the change of the potential (the other metric
much smaller) with time is very small (less then $O(2)$ level),
only the leading term is considered. Therefore the term
$\frac{2}{c^2} \int_A^B \frac{\partial w}{\partial t} dt$ is
already on $O(4)$ level, but not $O(2)$. Furthermore, if we take
Eq.(\ref{A7}) into account, we have $dx = dl + O(2) = c dt +
O(2)$. Then Eq.(\ref{A8}) simplifies to
\begin{equation}\label{A9}
\int_A^B \frac{\partial F}{\partial t} dx =
  \frac{2}{c^2} \int_A^B \frac{\partial w}{\partial t}dt
  + O(6) \, .
\end{equation}
If we consider quick variable field (e.g. field in pulsar) we have
to take Eq.(\ref{A8}) to substitute into Eq.(\ref{A9}).

\noindent
The second integral term of Eq.(\ref{25eq}) then becomes
\begin{equation}\label{59eq}
\frac{\sqrt{G_B}}{\sqrt{G_A}} \frac{2\eta}{\left( 1- F(B)
  \frac{{\bf k}_B \cdot {\bf v}_B}{|k_B|}\right)c^2}
  \int_A^B \frac{\partial w}{\partial t} dt \, .
\end{equation}
Gathering all evaluations done thus far in this section we arrive
at following general formula for the solar system:
\begin{equation}\label{60eq}
\frac{\Delta \tau_B}{\Delta \tau_A} = \sqrt{\frac{G_B}{G_A}}
 \frac{1}{\left( 1 - F(B) \frac{{\bf k}_B \cdot {\bf v}_B}{|k_B|}
 \right) }
 \left( 1 - F(A) \frac{{\bf k}_A \cdot {\bf v}_A}{|k_A|}
 + \frac{2 \eta}{c^2} \int_A^B \frac{\partial w}{\partial t}
 dt \right) \, .
\end{equation}
In Eq.(\ref{60eq}) we only consider the leading term, namely the
scalar potential changing with the time, but in our scheme all of
the second post Newtonian terms (see Eq.(\ref{A8})) can be
included in. Maybe in a system of binary pulsars, the higher order
terms in Eq.(\ref{A8}) are important.  For static metric and in
1PN level Formula (\ref{60eq}) agrees to the known formula in
\cite{adle75}.

We proceed by evaluating the remaining terms in (\ref{60eq}).
 From Eq. (\ref{51eq}) and Eq. (\ref{52eq}), we find
\begin{eqnarray}
\sqrt{G_B} &=& 1 - \frac{w(B)}{c^2}- \frac{v^2_B}{2c^2}
  + \frac{w^2(B)}{2c^4} + \frac{4w_i(B)v^i_B}{c^4}
  - \frac{3w(B)v^2_B}{2c^4} - \frac{v^4_B}{8c^4}
  \, , \label{61eq} \\
\frac{1}{\sqrt{G_A}} &=& 1 + \frac{w(A)}{c^2}
   + \frac{v^2_A}{2c^2} + \frac{w^2(A)}{2c^4}
   - \frac{4w_i(A)v^i_A}{c^4}
  + \frac{5w(A)v^2_A}{2c^4} +\frac{3v^4_A}{8c^4}
  \, . \label{62eq}
\end{eqnarray}
We also have
\begin{eqnarray}
\left( 1 - F(B) \frac{{\bf k}_B \cdot {\bf v}_B}{|k_B|}
  \right)^{-1}&=&1 + \frac{{\bf k}_B \cdot {\bf v}_B}{c|k_B|}
  + \frac{1}{c^2} \left( \frac{{\bf k}_B \cdot {\bf v}_B}
  {|k_B|} \right)^2
  +\frac{2w(B)}{c^3} \frac{{\bf k}_B \cdot {\bf v}_B}{|k_B|}
  +\frac{1}{c^3} \left( \frac{{\bf k}_B \cdot {\bf v}_B}
  {|k_B|} \right) ^3 \nonumber \\
  && -\frac{4w_i(B)}{c^4} \left. \frac{dx^i}{dx} \right| _B
  \frac{{\bf k}_B \cdot {\bf v}_B}{|k_B|}
  +\frac{4w(B)}{c^4} \left( \frac{{\bf k}_B \cdot {\bf v}_B}
  {|k_B|} \right) ^2
  +\frac{1}{c^4} \left( \frac{{\bf k}_B \cdot {\bf v}_B}
  {|k_B|} \right) ^4 \, . \label{62eq2}
\end{eqnarray}
Substituting Eqs.(\ref{54eq}), (\ref{54eq2}), (\ref{61eq}),
(\ref{62eq}) and (\ref{62eq2}) into Eq.(\ref{60eq}), we finally
have a general formula for the comparison of clock rates in the
solar system, on the 2PN level of precision:
\begin{eqnarray}
\frac{\Delta \tau_B}{\Delta \tau_A} &=& 1 +  \Biggl\{
  \frac{1}{c^2} \biggl( w(A)-w(B) \biggr)
  + \frac{1}{2c^2} \left( v^2_A - v^2_B \right)
  - \left( \frac{{\bf k}_A \cdot {\bf v}_A}{c|k_A|}
  - \frac{{\bf k}_B \cdot {\bf v}_B}{c|k_B|} \right) \nonumber \\
 && - \frac{1}{c^2}\left( \frac{{\bf k}_B \cdot {\bf v}_B}{|k_B|}
  \right) \left( \frac{{\bf k}_A \cdot {\bf v}_A}{c|k_A|}
  - \frac{{\bf k}_B \cdot {\bf v}_B}{c|k_B|} \right) \Biggr\}
  \nonumber \\
 && + \frac{1}{c^3} \Biggl\{ \biggl( w(B)- w(A) \biggr)
  \left( \frac{{\bf k}_B \cdot {\bf v}_B}{c|k_B|}
  + \frac{{\bf k}_A \cdot {\bf v}_A}{c|k_A|} \right)
  + 2w(A) \left( \frac{{\bf k}_B \cdot {\bf v}_B}{c|k_B|}
  - \frac{{\bf k}_A \cdot {\bf v}_A}{c|k_A|} \right) \nonumber \\
 && - \frac{1}{2} \biggl( v^2(B) - v^2(A) \biggr)
  \left( \frac{{\bf k}_B \cdot {\bf v}_B}{c|k_B|}
  - \frac{{\bf k}_A \cdot {\bf v}_A}{c|k_A|} \right)
  + \left( \frac{{\bf k}_B \cdot {\bf v}_B}{c|k_B|} \right) ^2
  \left( \frac{{\bf k}_B \cdot {\bf v}_B}{c|k_B|}
  - \frac{{\bf k}_A \cdot {\bf v}_A}{c|k_A|} \right) \Biggr\}
  \nonumber \\
 && +\frac{1}{c^4}\Biggl\{ \frac{1}{2}\biggl( w(B)-w(A)\biggr)^2
   + \frac{1}{2} \biggl( 10w(A)-w(B) \biggr) v_A^2
   - \frac{1}{2} \biggl( w(A)+ 6w(B) \biggr) v_B^2 \nonumber \\
 && + \frac{1}{8} \biggl( 3v_A^4 - 2v_A^2v_B^2 - v_B^4 \biggr)
  + 4 \Biggl( w_i(B)v_B^i - w_i(A)v_B^i
  - w_i(B) \frac{k_B^i}{|k_B|}
  \frac{{\bf k}_B \cdot {\bf v}_B}{c|k_B|} \nonumber \\
 && + w_i(A) \frac{k_A^i}{|k_A|}
  \frac{{\bf k}_A \cdot {\bf v}_A}{c|k_A|} \Biggr)
  + \left( \frac{{\bf k}_B \cdot {\bf v}_B}{c|k_B|} \right) ^2
  \biggl( 3w(B) + w(A) + \frac{1}{2}(v_A^2 - v_B^2) \biggr)
  \nonumber \\
 &&  - 2 \left( \frac{{\bf k}_B \cdot {\bf v}_B}{c|k_B|}\right)
   \left( \frac{{\bf k}_A \cdot {\bf v}_A}{c|k_A|} \right)
   \biggl( w(A) + w(B) \biggr)
   + \left( \frac{{\bf k}_B \cdot {\bf v}_B}{c|k_B|} \right) ^3
  \left( \frac{{\bf k}_B \cdot {\bf v}_B}{c|k_B|}
  - \frac{{\bf k}_A \cdot {\bf v}_A}{c|k_A|} \right) \Biggr\}
  \nonumber \\
  && + \frac{2\eta}{c^2}
  \int_A^B \frac{\partial w}{\partial t} dt
  + O(5) \, . \label{63eq}
\end{eqnarray}
Eq.(\ref{63eq}) gives the frequency shift from A to B to the order
$1/c^4$ . If we really want to apply this equation in the case of
the solar system, $w(A), \, w(B), \, w_i(A)$ and $w_i(B)$ have to
be expanded by means of internal and external multiple moments
\cite{damo91}.  As the calculation is rather long, we may publish
it in a separate paper. We have checked that the leading terms of
Ref.\cite{line02} are included in the expansion of
Eq.(\ref{63eq}).

To sum up our discussion we can say that Formula (\ref{63eq})
`contains' the Doppler effect, transverse Doppler effect
(relativistic Doppler effect), gravitational redshift and their
complete coupling effects to 2PN level in the  solar system. In
addition there is a term which is the integral of the rates of
change of the scalar potential along the null geodetic line from
source A to receiver B. This is probably the most interesting
result in our paper. Hopefully this integral term and the coupling
effects can be tested in the future with a deep space explorer and
are confirmed.

%%%%%%%%%%%%%%%%%%%%%%%%%%%%%%%%%%%%%%%%%%%%%%%%%%%%%%%%%%%%%%%%%%%%%%%%

\section{Conclusions}

\qquad 1. We have synthesized all known effects for the comparison
of clock rates in one formula Eq.(\ref{25eq}). The synthesized
formula contains additional coupling terms and a new integral
terms and thus gives essential new but untested information.
Therefore, this synthesized formula is not an end in itself, but a
starting point for the further test work. We hope that, this work
could contribute to the further comparison of clock rates, such as
ACES mission planned in near future.

2. The general form is valid for any metric gravitational theory,
not only for general relativity. If we substitute the parametrized
2PN metric into the formula, Eq.(\ref{63eq}) could include
parameters.

3. In the case of cosmology, the general form can be used for any
linearly perturbed metric, in particular it allows to include the
general Sachs-Wolfe effects.

4. In section IV, we have discussed the comparison of two clock
rates both on the earth and a space station with 2PN precision. In
fact, the calculation of the higher precision might be done in a
similar way, if we know the metric to higher order.

5. Eq.(\ref{25eq}) or  Eq.(\ref{60eq}) say that the clock rates
depend on the trajectory of the transmitted signal
 and on the metric (especially the scalar potential)
varying with time. While (\ref{25eq}) is a general formula,
Eq.(\ref{60eq}) is valid only in the solar system; however if
replace Eq.(\ref{A9}) by the integral of Eq.(\ref{A8}), a general
2PN formula results.

6. The general form may be taken as the basis for a starting point
to compare clock rates at any two different space time points. For
example, the frequency shift caused by gravito-magnetic effect (or
Lens-Thirring effect) can also be considered in our scheme.

%%%%%%%%%%%%%%%%%%%%%%%%%%%%%%%%%%%%%%%%%%%%%%%%%%%%%%%%%%%%

\vspace{0.6cm}
\noindent{\bf Acknowledgments}

\bigskip\noindent
This work was supported by the National Natural Science Foundation
of China (Grant No. 10273008). In the workshop ``Relativistic
Astrophysics and High-precision Astrodynamics" led by Prof.
Wei-tou Ni, we had several very fruitful discussions, and
especially we would like to thank Prof. Tianyi Huang for his
useful discussion. We also would like to thank the referees for
their comments and pointing out and correcting our mistakes.

%\vspace*{0.5cm}

%%%%%%%%%%%%%%%%%%%%%%%%%%%%%%%%%%%%%%%%%%%

\end{document}